\documentclass[bimj,fleqn]{w-art}
\usepackage{times}
\usepackage{w-thm}
\usepackage[authoryear]{natbib}
\theoremstyle{plain}

\theoremstyle{definition}

\usepackage[]{graphicx}
\chardef\bslash=`\\ 

\usepackage{booktabs}  
\usepackage{verbatim}  
\usepackage{hyperref,breakurl}  
\usepackage{booktabs}  
\usepackage{multirow}  %
\usepackage{tcolorbox}  
\tcbuselibrary{breakable}

\providecommand{\normaldistn}{\mathrm{Normal}}
\providecommand{\halfnormaldistn}{\mathrm{half\mbox{-}Normal}}
\providecommand{\prob}{\mathrm{P}}
\providecommand{\unifdistn}{\mathrm{Uniform}}

\begin{document}
\DOIsuffix{bimj.DOIsuffix}
\Volume{XX}
\Issue{YY}
\Year{2023}
\pagespan{1}{}

\title{Investigating the heterogeneity of ``study twins''}
\author[C.~R\"{o}ver]{Christian R\"{o}ver\footnote{Corresponding author: {\sf{e-mail: christian.roever@med.uni-goettingen.de}}}\inst{,1}} 
\address[\inst{1}]{Department of Medical Statistics, University Medical Center G\"{o}ttingen, Humboldtallee~32, 37073~G\"{o}ttingen, Germany}
\address[\inst{2}]{DZHK (German Center for Cardiovascular Research, partner site G\"{o}ttingen, 37073~G\"{o}ttingen, Germany}
\author[T.~Friede]{Tim Friede\inst{1,2}}
\Receiveddate{xxx} \Reviseddate{yyy} \Accepteddate{zzz} 


\begin{abstract}
  Meta-analyses are commonly performed based on random-effects models, while in certain cases one might also argue in favour of a common-effect model. One such case may be given by the example of two ``study twins'' that are performed according to a common (or at least very similar) protocol.  
  Here we investigate the particular case of meta-analysis of a pair of studies, e.g.\ summarizing the results of two confirmatory clinical trials in phase~III of a clinical development programme. Thereby, we focus on the question of to what extent homogeneity or heterogeneity may be discernible, and include an empirical investigation of published (``twin'') pairs of studies.
  A pair of estimates from two studies only provides very little evidence on homogeneity or heterogeneity of effects, and ad-hoc decision criteria may often be misleading.
\end{abstract}

\maketitle

\section{Introduction}
  In drug licensing, evidence from several independent sources is often required to demonstrate efficacy and safety of novel treatments \citep{EMEA2001,KennedyShaffer2017}.
  Especially in health technology assessment (HTA), meta-analysis methods are commonly applied to combine and judge evidence from several sources, e.g. typically a small number of clinical trials conducted in the later development phases \citep{DiasEtAl2013a,BenderEtAl2018}.

  The case of meta-analysis of only few studies, however, presents particular challenges. The assessment of heterogeneity (or homogeneity) of effect estimates turns out central in the evaluation of implications \citep{Cochran1937,VeronikiEtAl2016,FriedeRoeverWandelNeuenschwander2017a,FriedeRoeverWandelNeuenschwander2017b}.
  In the case of $k\!=\!2$~studies, \citet{BenderEtAl2018} proposed to assess plausibility of the homogeneity assumption based on contextual information, and if possible resort to a common-effect analysis.  A common-effect analysis would effectively mean a relatively simple ``pooling'' of effect estimates, while \emph{stratification} by study is usually still implemented \citep{Gail2005}. In a certain sense, however, homogeneity (or zero heterogeneity) may also be considered a ``most optimistic'' assumption \citep{RoeverEtAl2021}, so it may not be a suitable ``default'' option.

  As a prime example where homogeneity of treatment effects may often be assumed, \emph{``twin studies''} are sometimes quoted,  that is, when a common study protocol is implemented for a pair of otherwise independently conducted trials \citep{BenderEtAl2018,SchulzEtAl2022}. The reasoning here is that with similar study populations as well as treatment and outcome assessment procedures, one may expect the eventual quantitative outcome also to be consistent \citep[Sec.~10.10]{CochraneHandbook}. This might be particularly true for relative effect measures, which may be able to provide rather homogeneous effect estimates despite heterogeneous study populations.
  The degree to which outcomes are still subject to between-study heterogeneity may depend on a range of factors, such as the medical area \citep{RhodesEtAl2015,TurnerEtAl2015}, the endpoint considered, or type of effect measure \citep{DeeksAltman2001}. For example, an odds ratio based on an overall mortality endpoint might be determined rather unambiguously, while treatment effects of a psychological intervention may be harder to quantify consistently.

  The recent example of the \textsc{Respire} trials highlights how even the suggestion of the potential presence of heterogeneity may generate considerable unease \citep{ChotirmallChalmers2018}.
  However, judging the presence or magnitude of heterogeneity among a pair of studies is very hard; heterogeneity would need to be substantial in order to become apparent, and often rather ad-hoc criteria are consulted, which may in fact not constitute very conclusive evidence.
  Given the commonly substantial uncertainty around the homogeneity assumtion, and the potential consequences of its violation, one might then prefer to anticipate the possibility of heterogeneity in the analysis.

  Here we investigate the interface between a common-effect and a random-effects meta-analysis of a pair of ``study twins'', with a focus on the question to what extent one can actually judge whether heterogeneity is present or absent in a pair of estimates.
  We will also illustrate heterogeneity issues by considering an empirical sample of pairs of published study twins.

  In the following, we will first introduce some context and terminology; we will then consider examples of \emph{study twins} in detail. We will investigate to what extent heterogeneity can be recognized from only a pair of studies, and we will consult an empirical sample of study twins for commonly observed (empirical) heterogeneity. We will eventually return to the examples and close with some conclusions.

\section{Meta-analysis of few studies}
\subsection{Study twins}
  Regulatory guidelines commonly require independent replication of experimental results before these are considered convincing \citep{EMEA2001,KennedyShaffer2017}.
  The degree to which it is possible to replicate the results of one study in another study is also referred to as \emph{``(results) reproducibility''} \citep{GoodmanFanelliIoannidis2016,Plesser2018}.
  In order to try and generate consistent results in replicate studies, factors that might induce between-study heterogeneity (e.g., inclusion criteria, treatment regimens, outcome assessment, modeling assumptions, \ldots) are usually carefully controlled.
  \citet[Sec.~10.10]{CochraneHandbook} in this context distinguish between \emph{clinical} and \emph{methodological diversity}, i.e., heterogeneity in study populations and design aspects, which may eventually be reflected in the \emph{statistical heterogeneity}, i.e., the heterogeneity in the eventual quantitative study outcomes.
  In this spirit, \citet{BenderEtAl2018} characterized \emph{twin studies} as a pair of trials ``\emph{where the same (or at least a very similar) study protocol is replicated in a second study}''.
  For the present purpose, we define \emph{study twins} as a study pair that is: (i)~either jointly planned, or where one is planned as a replication of the other; and (ii)~(essentially) identical especially with respect to inclusion criteria and treatment details.

\subsection{Between-trial heterogeneity in treatment effects}
  Effect estimates from clinical trials are typically accompanied by measures of uncertainty, e.g., standard errors or confidence intervals. \emph{Homogeneity} of different estimates then means that, despite numerical inequality, these refer to a common underlying ``true'' value. Quite commonly, however, there is additional variability present going beyond what could be attributed to mere measurement error; such excess variance is termed \emph{heterogeneity}. Unless such heterogeneity can be attributed to certain trial-specific features (e.g., in a meta-regression), it is commonly accounted for by including extra variance components (random effects) into the analysis models \citep{Cochran1937,AdesLuHiggins2005,HigginsThompsonSpiegelhalter2009}.

  It is important to note that assuming treatment effect homogeneity does not necessarily lead to a ``na\"{i}ve'' pooling of data. When the estimates to be combined themselves result as differences or contrasts (e.g., as an odds ratio comparing a treatment to a control group), then the ``common-effect'' analysis (e.g., referring to a common odds ratio among several experiments) will usually correspond to a \emph{stratified} analysis allowing for differences e.g.\ between the individual studies' control groups \citep{Gail2005,DeeksAltman2001}.

\subsection{The normal-normal hierarchical model}
  In the following, we will consider the normal-normal hierarchical model (NNHM), a simple random-effects model that constitutes the basis of many meta-analyses, and where homogeneity and heterogeneity simply correspond to zero or positive values for the heterogeneity variance component \citep{Fleiss1993,AdesLuHiggins2005,HigginsThompsonSpiegelhalter2009,Roever2020}. Within the NNHM, estimates~$y_i$ and their associated standard errors~$s_i$ (where $i=1,\ldots,k$) are modeled as
  \begin{equation}
    y_i | \theta_i      \;\sim\; \normaldistn(\theta_i, s_i^2) \quad \mbox{and} \quad
    \theta_i | \mu,\tau \;\sim\; \normaldistn(\mu, \tau^2)
  \end{equation}
  leading to the marginal expression
  \begin{eqnarray}
    y_i | \mu,\tau & \sim & \normaldistn(\mu, \tau^2 + s_i^2)
  \end{eqnarray}
  \citep{Fleiss1993,AdesLuHiggins2005,HigginsThompsonSpiegelhalter2009,Roever2020}.
  Homogeneity ($\tau=0$) leads to the special case of the \emph{common-effect} model (with $\theta_i=\mu$), and in case of heterogeneity ($\tau>0$), the estimates~$y_i$ relate to ``study-specific'' mean parameters~$\theta_i$ that are not identical, but only similar across studies.

  Assessment of heterogeneity is challenging in case of few studies (small~$k$), since significance tests suffer from low power, quantification is sensitive to the choice of estimator \citep{HardyThompson1998,GavaghanEtAl2000,Ioannidis2008,PereiraEtAl2010,VeronikiEtAl2016}, and estimation uncertainty is hard to assess \citep{VeronikiEtAl2019}.
  More sophisticated (frequentist) confidence intervals also incorporating the uncertainty originating from heterogeneity estimation have been proposed, e.g., the Hartung-Knapp-Sidik-Jonkman (HKSJ) or modified Knapp-Hartung (mKH) intervals \citep{RoeverKnappFriede2015}.
  These address the issues to some degree, while some concerns regarding their general appropriateness remain \citep{JacksonEtAl2017}, and the actual performance (in particular, interval widths) for as few as two studies may still be unsatisfactory \citep{RoeverKnappFriede2015,FriedeRoeverWandelNeuenschwander2017b}.
  In a Bayesian context, the likelihood conveys little information on the heterogeneity, and the prior tends to remain influential for inference \citep{Roever2020,RoeverEtAl2021}.
  While for Bayesian methods this is no fundamental obstactle, prior specification then may require particularly solid motivation \citep{FriedeRoeverWandelNeuenschwander2017a,FriedeRoeverWandelNeuenschwander2017b,RoeverEtAl2021}.

\section{Two examples}
\subsection{The GLOW trials}\label{sec:glow1}
  The \textsc{Glow~1} and \textsc{Glow~2} studies are randomized controlled trials (RCTs) that investigated the use of glycopyrronium bromide (50$\mu$g daily) in patients with moderate to severe chronic obstructive pulmonary disease (COPD) \citep{DUrzoEtAl2011,KerwinEtAl2012}.  Both were phase~III studies using the same set of inclusion criteria; only the \textsc{Glow~2} study was of a longer duration and included another additional treatment arm.
  A pooled analysis was also published subsequent to the individual studies' results \citep{DUrzoEtAl2013}.

  \begin{figure}[t]
    \centering
    {\includegraphics[width=0.8\linewidth]{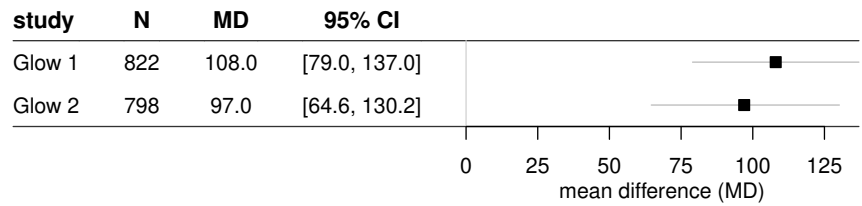}}
    \caption{\label{fig:glow} The \textsc{Glow~1/2} example
      data.  The effect measure is the mean difference (MD) in \emph{trough FEV\textsubscript{1} at week~12} compared to placebo (mL).  Both studies show consistent results.}
  \end{figure}

  The primary endpoint related to the \emph{forced expiratory volume in 1~second (FEV\textsubscript{1})}.  The \emph{trough FEV\textsubscript{1}} is defined as the ``mean of the values at 23~h 15~min and 23~h 45~min after dosing''.  The treatment effects (mean differences in \emph{trough FEV\textsubscript{1} at week~12}, relative to placebo) are shown in Figure~\ref{fig:glow}.  A positive difference here indicates a benefit for the patient.  With an average increase in volume by roughly 100~mL in both cases, the studies showed consistent results.

\subsection{The RESPIRE trials}\label{sec:respire1}
  Recently, results of the \textsc{Respire~1} and \textsc{Respire~2} studies were published, which were two randomized controlled trials investigating cipro\-floxacin for the treatment of non-cystic fibrosis bronchiectasis \citep{DeSoyzaEtAl2018,AksamitEtAl2018}.  
  The two studies may be considered \emph{twins}, yet the outcomes turned out somewhat discordant, triggering lively dissussions of the role of heterogeneity \citep{AksamitEtAl2018,ChotirmallChalmers2018}.

\begin{figure}[t]
    \centering
    {\includegraphics[width=0.9\linewidth]{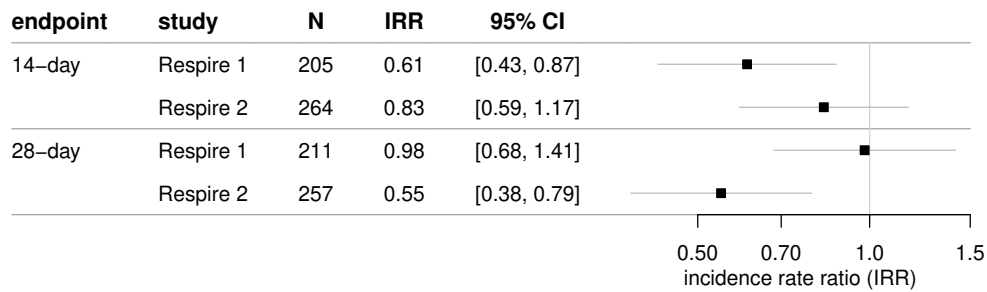}}
    \caption{\label{fig:respire} The \textsc{Respire~1/2} example
      data.  The treatment effect here is expressed in terms of incidence rate ratios (IRRs) for exacerbation events. Not only
      do the effect estimates differ between trials, but also the
      overall picture differs for the two (14-day and 28-day)
      endpoints.}
  \end{figure}

  The \textsc{Respire~1} and \textsc{Respire~2} studies included 416 and 521~patients, respectively. Each study had four treatment groups; two treatment regimens (14-day vs.\ 28-day on/off cycles for 48 weeks) were each compared to a corresponding control group.  
  Figure~\ref{fig:respire} shows both studies' data on the primary endpoint \emph{incidence rate ratio (IRR) of exacerbations} for the two regimens (14-day and 28-day).
  
  In the 14-day dosing regimen, the CIs are overlapping and mutually covering the other study's estimate.  In the 28-day regimen, CIs are also overlapping, but do not include each other's estimates. In both cases, there is one significant and one non-significant outcome; however, in the 14-day regimen, only \textsc{Respire~1} showed a significant effect, while in the 28-day regimen, \textsc{Respire~2} turned out significant.

  The evidence from the two studies appears somewhat puzzling: each indicates significance for one of the treatment regimens, while none of the two is consistently superior. This discordant pattern of estimates probably added to the controversy here. There also was a second co-primary survival (hazard ratio) endpoint, which similarly did not point to a clear-cut conclusion.

  A lively discussion of results and of potential sources of heterogeneity followed \citep{AksamitEtAl2018,ChotirmallChalmers2018}. While the two studies had identical inclusion and exclusion criteria, they had been conducted in different geographical regions, and differing COPD prevalences as well as (significantly) differing exacerbation rates at baseline were pointed out.
  \citet{ChotirmallChalmers2018} note a \emph{``failure of replicate trials to reproduce results''} here.

\section{Evidence on heterogeneity from a pair of studies}
  In the case of a single pair of studies ($k=2$), homogeneity or heterogeneity empirically only manifests itself in the difference between the two estimates ($y_1$ and $y_2$). 
  The two corresponding 95\% confidence intervals (CIs) are calculated as $y_i \pm z_{0.975} s_i$, where $z_{0.975}\approx 1.96$ is the 97.5\%~quantile of the standard normal distribution.
  When looking at such a pair of estimates along with their associated CIs, several features may suggest homogeneity:
  \begin{itemize}
    \item overlapping CIs, 
    \item a non-significant test for homogeneity (e.g., Cochran's $Q$~test),
    \item CIs mutually covering the other study's effect estimate,
    \item a zero heterogeneity estimate ($\hat{\tau}=0$).
  \end{itemize}
  In the case of two studies, the evidence to decide between homogeneity and heterogeneity commonly is very weak. Significance tests for homogeneity generally are known to have little power, especially for as few as $k=2$ studies \citep{HardyThompson1998,Ioannidis2008,PereiraEtAl2010}, and heterogeneity often tends to be underestimated \citep{KontopantelisSpringateReeves2013}. For some more details on Cochran's $Q$~test, see also Appendix~\ref{sec:CochranQ}.

  For the simple case of two studies, the probabilities for the above four events may be computed numerically (see the appendix for details).  Figure~\ref{fig:probabilities} illustrates these for the special case of equal standard errors ($\sigma_1=\sigma_2=\sigma$, so that the $I^2$~heterogeneity measure results as $I^2=\frac{1}{1+(\frac{\sigma}{\tau})^2}$ and only depends on the relative magnitudes of~$\tau$ and~$\sigma$ \citep{HigginsThompson2002}).
  \begin{figure}[t]  
    \centering
    \makebox{\includegraphics[width=0.95\linewidth]{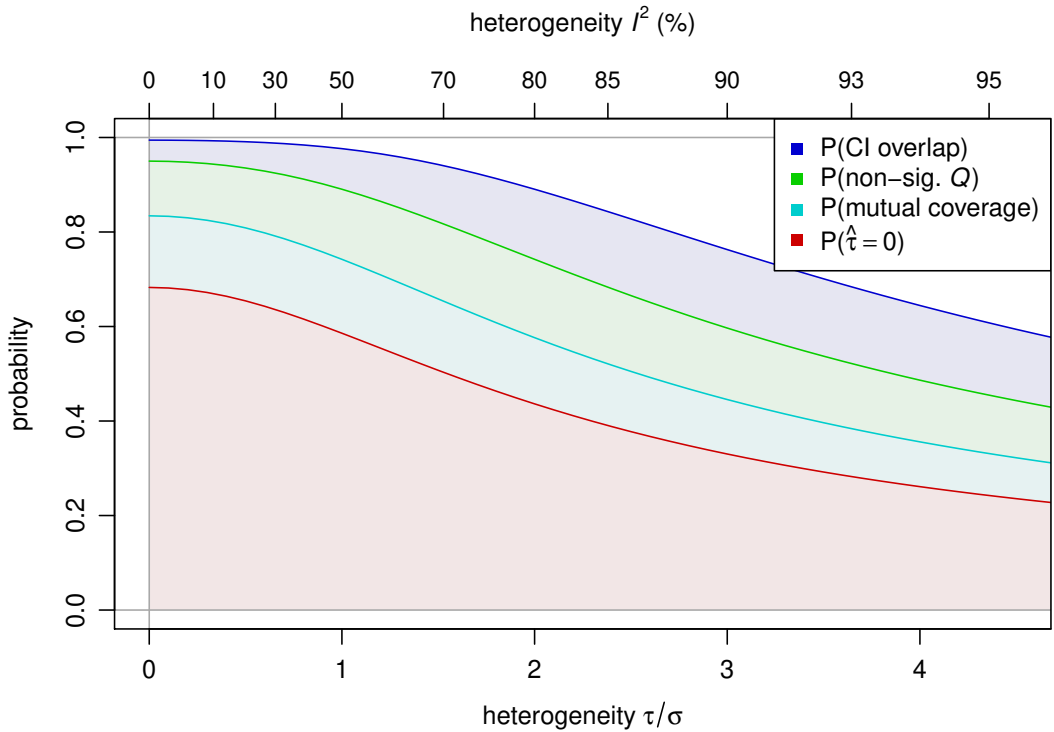}}
    \caption{\label{fig:probabilities} Probabilities of overlapping
      CIs, a non-significant $Q$-test, CIs with mutually covered
      effect estimates, or a zero heterogeneity estimate for the
      special case of equal standard errors
      ($\sigma_2=\sigma_1=\sigma$). In this case, all these
      probabilities only depend on the ratio of~$\frac{\tau}{\sigma}$
      (and, with that, on~$I^2$).}
  \end{figure}
  Note that all of these (CI overlap, a non-significant $Q$-test, mutually covered effect estimates, or a zero heterogeneity estimate) are often taken to be indicative of homogeneity. From the figure one can see that all of these are likely to occur even for substantial amounts of heterogeneity, i.e., the data may in that sense often falsely suggest homogeneity, or may at least ``look homogeneous''.
  The reverse is also not uncommon: under homogeneity, the data may also exhibit features suggestive of some heterogeneity with non-negligible probability.
  Consequently, homogeneity and heterogeneity are hard to distinguish based only on a pair of estimates.

\section{An empirical investigation}
  \subsection{Aim, search strategy and data extraction}
    To gain some insights into homogeneity or heterogeneity among published pairs of studies that one may consider to be ``study twins'', we performed a literature search to gather some examples.
    We devised a relatively simple search strategy in order to obtain a sizeable number of studies, taking advantage of the meanwhile common requirement for pre-registration of clinical trials, e.g., in the United States National Library of Medicine (NLM) registry (``\texttt{clinicaltrials.gov}'') \citep{DeAngelisEtAl2005}.
    We focused on publications in the \emph{New England Journal of Medicine (NEJM)} in the decade of 2010--2019. Due to the NEJM's policy of quoting \texttt{clinicaltrials.gov} identifiers in the abstract, we were able to single out studies referring to more than one identifier based on an exported list of abstracts.
    In order to select pairs of ``study twins'' among the abstracts in question, we then employed the following inclusion criteria:
    \begin{itemize}
      \item studies have been planned jointly
      \item studies are assigned to the same clinical phase
      \item differences like additional arms, differing follow-up times, or additional investigations are unproblematic as long as the primary endpoint is unaffected.
    \end{itemize}
    Exclusion criteria were:
    \begin{itemize}
      \item differences in the studies' inclusion criteria (age groups, diseases, severities, genotypes, \ldots)
      \item differing treatments, doses, or modes of administration (e.g., subcutaneous vs. intravenous).
    \end{itemize}
    We then focused on the primary endpoint; in case of several co-primaries, we picked the one reported first; in case no primary endpoint is explicitly specified, we picked the first endpoint quoted in the ``results'' section.
    From the qualifying studies, we eventually extracted the following information:
    \begin{itemize}
      \item publication reference (DOI),
      \item study names (or acronyms), NCT identifiers,
      \item primary endpoint name and effect measure,
      \item whether primary endpoint analyses were pooled or reported separately,
      \item primary endpoint data (estimates and standard errors). 
    \end{itemize}

  \subsection{Search results}
    A PubMed search for articles published in the decade 2010--2019 in NEJM and quoting the term ``clinicaltrials.gov'' in the abstract yielded 1371~references. Out of these, 103 quoted more than one \texttt{clinicaltrials.gov} (NCT) identifier. 
    30~studies met the inclusion criteria, and 26~studies reported adequate data, i.e., eventually yielding 26~twin pairs of studies.

    \begin{figure}[t]  
      \centering
      {\includegraphics[width=0.7\linewidth]{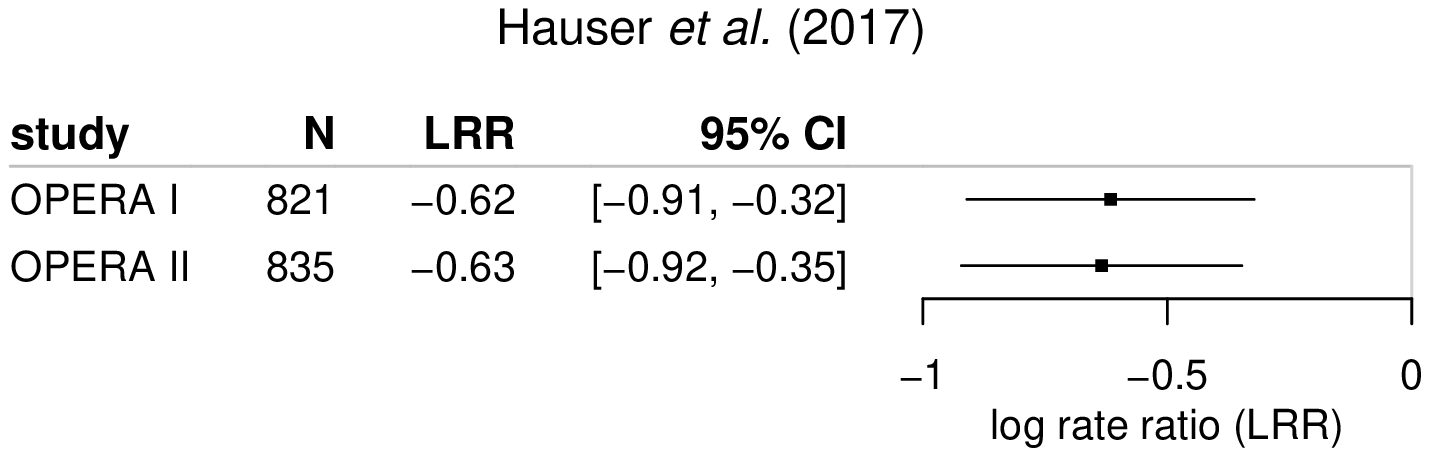}\\ 
               \includegraphics[width=0.7\linewidth]{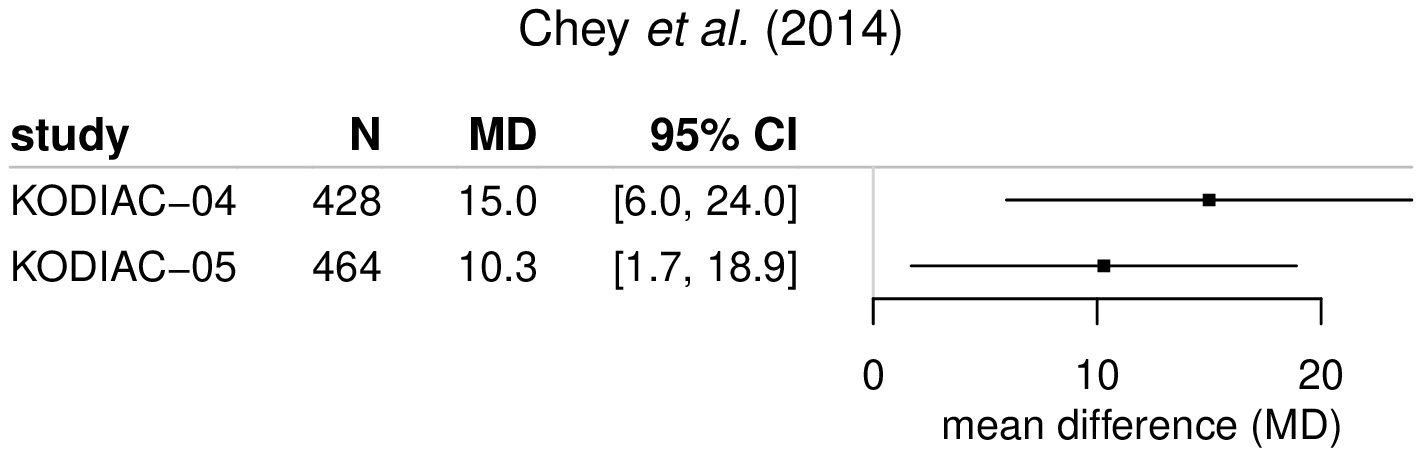}\\
               \includegraphics[width=0.7\linewidth]{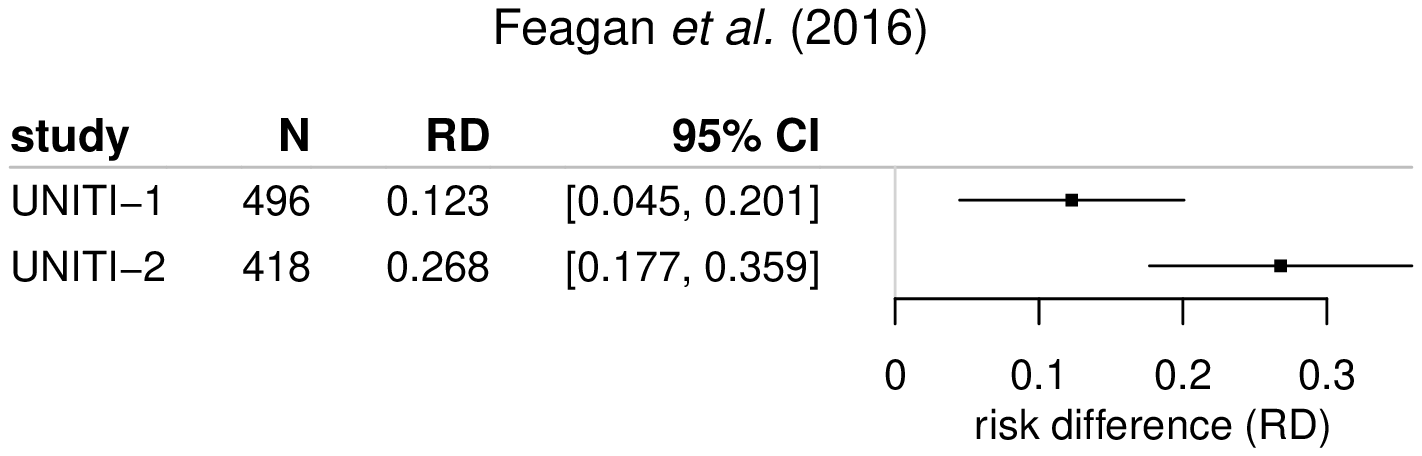}}
      \caption{\label{fig:forest} Three examplary forest plots
        corresponding to ``small'', ``moderate'' and ``large'' values for
        Cochran's $Q$~statistic within the data set
        \citep{HauserEtAl2017,CheyEtAl2014,FeaganEtAl2016}.}
    \end{figure}

    Figure~\ref{fig:forest} illustrates three example cases that were selected based on their associated Cochran's $Q$~statistic values (smallest, largest, and a moderate one); the ones shown here then correspond to
    $Q=0.0076$ ($p=0.93$),
    $Q=0.54$ ($p=0.46$) and
    $Q=5.6$ ($p=0.017$), respectively.
    In all three cases, the two studies' CIs overlap; in the third case, the CIs however do not include the other study's estimate. Also, only in the third case heterogeneity is significant ($p=0.017$ for Cochran's $Q$~test), and the heterogeneity estimate is positive ($\hat{\tau}=0.093$). The complete data are available in the online supplement.

  \subsection{Descriptive analysis}\label{sec:descriptive}

    \begin{table} 
      \caption{\label{tab:frequencies01}The empirically observed frequencies in comparison to the theoretically expected probabilities (in percent) for selected heterogeneity values~$\tau/\sigma$ (see also Figure~\ref{fig:probabilities}).}  
      \centering
      \begin{tabular}{lcccccc}
        \toprule
          &      & & \multicolumn{4}{c}{event}\\
        \cmidrule(lr){4-7}
        & $\frac{\tau}{\sigma}$ & $I^2$& overlap & nonsig.\ $Q$ & mutual & $\hat{\tau}=0$ \\
        \midrule
        empirical frequency     &     & & 100.0 & 84.6 & 80.8 & 65.4 \\
        \cmidrule(lr){1-7}
                               & 0.0 & \phantom{0}0\% & \phantom{1}99.4 & 95.0 & 83.4 & 68.3 \\
        theoretical probability & 0.5 & 20\% & \phantom{1}99.1 & 93.5 & 80.9 & 65.4 \\
        (assuming $\sigma_1=\sigma_2$)   & 1.0 & 50\%  & \phantom{1}97.6 & 89.0 & 74.2 & 58.6 \\
                              & 2.0 & 80\% & \phantom{1}89.0 & 74.2 & 57.6 & 43.6 \\
        \bottomrule
      \end{tabular}
    \end{table}

    Table~\ref{tab:frequencies01} shows the empirically observed frequencies of CI overlap, a non-significant $Q$-test, mutual effect coverage of CIs, or a zero heterogeneity estimate for the empirical data.  Under homogeneity ($\tau=0$) and equal standard errors ($\sigma_1=\sigma_2$), the corresponding probabilities would be at 99.4\%, 95.0\%, 83.4\%, and 68.3\%, respectively (see also Figure~\ref{fig:probabilities}). Event probabilities are also shown for a selection of heterogeneity values $\tau/\sigma > 0$ (and assuming equal standard errors $\sigma_2=\sigma_1=\sigma$).  
    In order to match probabilities to the observed frequencies above, it would require heterogeneity values $\tau / \sigma$ of 1.3, 0.51 and 0.50, respectively.

    The study pairs were usually of roughly equal size and precision.
    The larger study had a median of a 1.04-fold sample size compared to the smaller study, and the largest ratio was at~2.78.
    The associated standard errors were also very similar: the larger standard error was a median of a 1.07~times greater than the smaller standard error; the maximum ratio was at a factor of~1.50.

    Table~\ref{tab:frequencies02} shows frequencies of combinations of effect estimates. In the majority of cases, both estimates were significant and were pointing in the same direction (where ``significant'' here means that $|\frac{y_i}{\sigma_i}|>1.96$).
    In only two cases, effect estimates were of opposing signs, and then both were also not significant.
    \citet{YusufCollinsPeto1984} also termed such cases of differing treatment effects with differing or equal signs as \emph{qualitative} and \emph{quantitative interactions}, respectively.

      \begin{table} 
        \caption{\label{tab:frequencies02}Frequencies of concordant and discordant effect estimates and significances among the 26~pairs of study twins.}  
        \centering
        \begin{tabular}{lccc}
          \toprule
          \multirow{2}*{\parbox[t][3.5ex][b]{8ex}{direction \\ of effect}} $\quad$  & \multicolumn{3}{c}{significance}\\
          \cmidrule(lr){2-4}
            & none & one & both \\
          \midrule
          $\;$ same     & 2 & 3 & 19 \\[0.5ex]
          $\;$ opposite & 2 & 0 &  \phantom{1}0 \\
          \bottomrule
        \end{tabular}
      \end{table}

    \begin{figure}[t]  
      \centering
      \makebox{\includegraphics[width=0.45\linewidth]{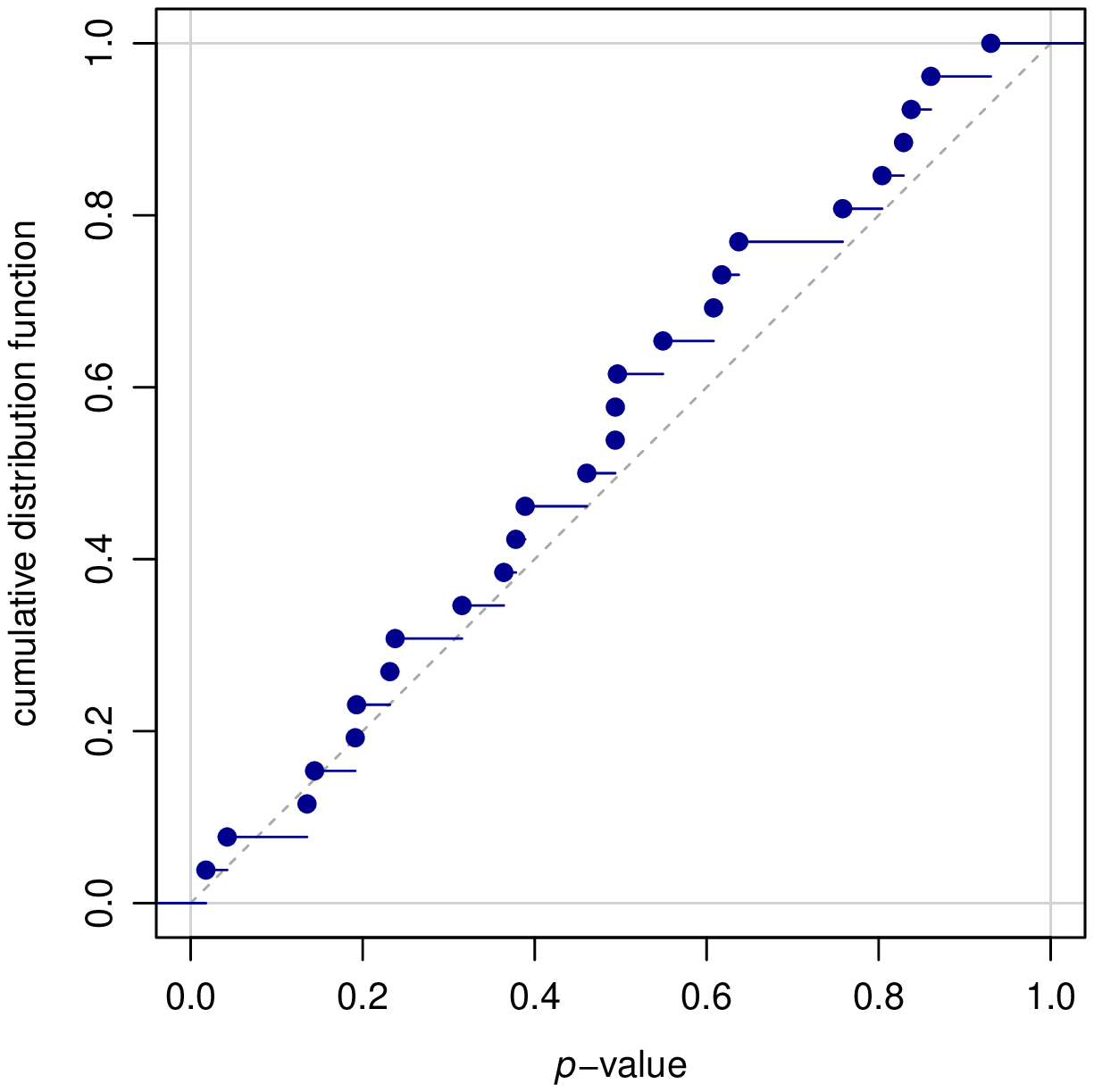}
               \hspace{0.08\linewidth}
               \includegraphics[width=0.45\linewidth]{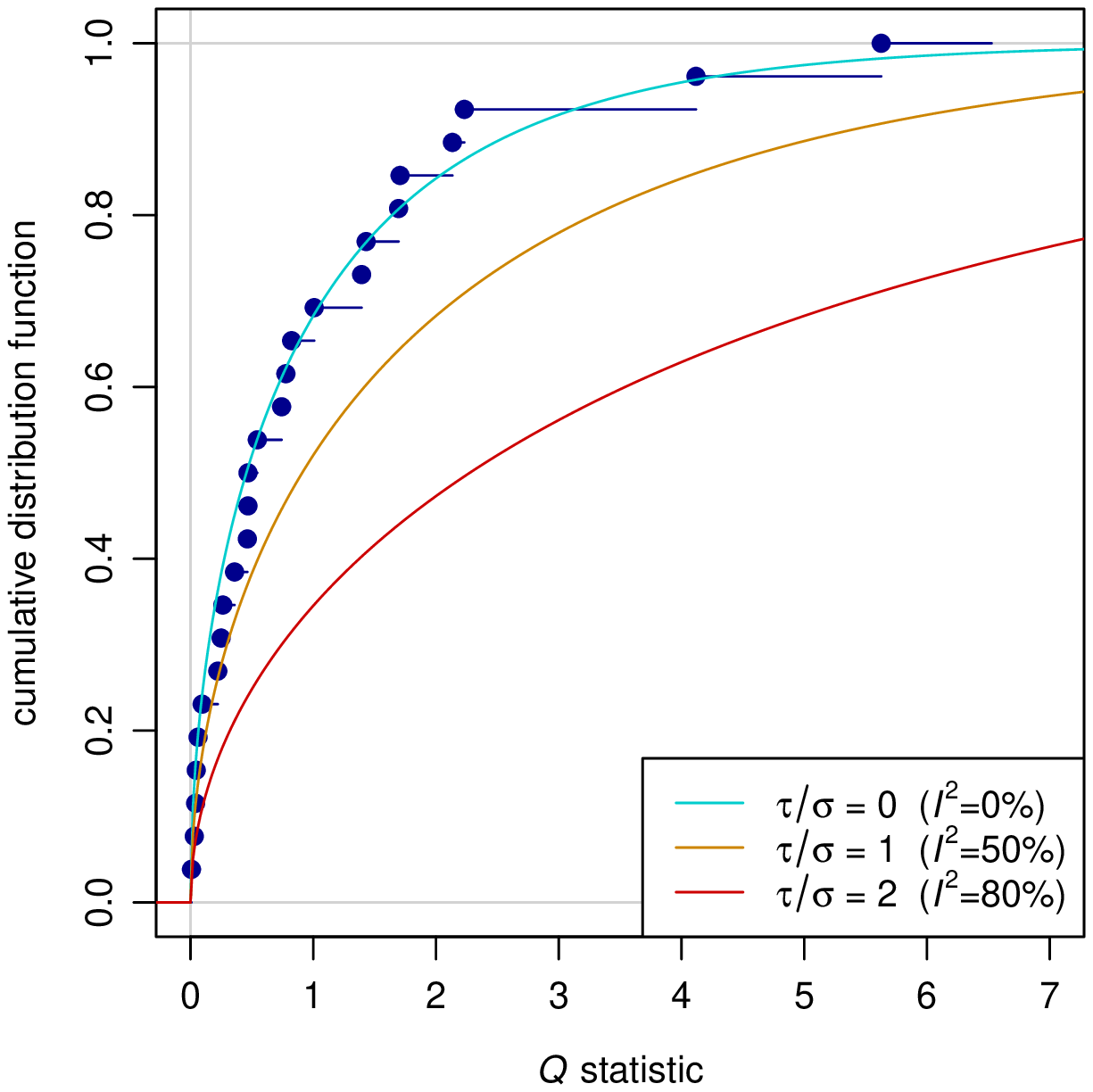}}
      \caption{\label{fig:pQvalues} Empirical cumulative distribution function (CDF) of the $Q$-test's $p$-values (left panel). The empirical distribution is close to a uniform distribution, which is indicated by the dashed line. The right panel shows the corresponding empirical distribution of test statistic ($Q$) values, which under the null hypothesis of homogeneity follow a $\chi^2_1$-distribution (bright blue line).}
    \end{figure}
    Figure~\ref{fig:pQvalues} (left panel) shows the empirical distribution of the $p$-values of Cochran's $Q$-test for heterogeneity. In case of homogeneity, these should follow a uniform distribution, as indicated by the dashed line.  A Kolmogorow-Smirnow-test for a uniform distribution of $p$-values yields $p=0.71$, i.e., no indication of non-uniformity.
    We may also investigate whether the lower tail of the distribution of $p$-values suggests non-uniformity; the minimum of $n$~uniformly distributed $p$-values follows a $\mathrm{Beta}(1,n)$-distribution.  The smallest $p$-value in the sample is at $p_{(1)}=0.0177$, which is at the $37\%$-quantile of the corresponding ($\mathrm{Beta}(1,26)$) distribution; it is hence also in no way ``unusually small'' under the hypothesis of homogeneity.
    
    The right panel of Figure~\ref{fig:pQvalues} shows a similar picture, but at the scale of the $Q$-test statistics (where large $Q$-values correspond to small $p$-values).
    Under the null hypothesis of homogeneity, these are $\chi_1^2$-distributed, as indicated by the bright blue line, and again the empirical distribution appears to match well. At the $Q$~scale, it is also possible to derive the distribution under certain alternatives (see the Appendix for details); two cases (with equal $\sigma_1=\sigma_2=\sigma$) are shown in the same plot.

  \subsection{Estimating heterogeneity}
    \subsubsection{Joint meta-analysis} 
      Heterogeneity values or their estimates~($\tau$ or~$\hat{\tau}$) for different meta-analyses are only comparable when the treatment effects are on the same or a similar scale \citep{RoeverEtAl2021}.  Among our sample of 26~study twins, the largest group with a common effect measure are 8~study pairs with log-OR as effect measure and 5~more studies with log-HR, log-rate-ratio or log-risk-ratio; these may be considered to relate to a common (logarithmic) scale. In the following, we will investigate these 13~twin pairs more closely by fitting variations of the NNHM to the data set.

    \subsubsection{Common heterogeneity parameter} 
      One way to quantify the heterogeneity (in particular in view of the possibility of homogeneity, i.e., $\tau=0$ for all 13~meta-analyses) is to fit a model with a single, \emph{common} heterogeneity parameter. Technically this may be implemented using a meta-regression approach, where pairs of (twin) studies are identified by a common indicator variable. Such a model may be fitted using either a frequentist or Bayesian approach, e.g., using the \texttt{metafor} or \texttt{bayesmeta} \textsf{R}~packages \citep{Viechtbauer2010,RoeverFriede2023}.

      A frequentist meta-analysis yields a (Paule-Mandel) point estimate of~$\hat{\tau}=0.0078$ and a 95\% confidence interval [$0.00$, $0.16$]. 
      A (Cochran's~$Q$) test for Heterogeneity turns out not significant ($p=0.42$).
      A Bayesian analysis (for both an improper uniform or a $\halfnormaldistn(0.5)$ prior) yields a posterior median of~$0.046$ along with a 95\% credible interval [$0.00$, $0.13$].
      When using proper, weakly informative priors ($\normaldistn(0,2.82)$ for the overall mean effect and $\halfnormaldistn(0.5)$ for the heterogeneity) \citep{GunhanRoeverFriede2020,RoeverEtAl2021}), a Bayes factor of~4 in favour of $\tau=0$ (homogeneity) may be computed \citep{KassRaftery1995,BDA3rd}.
      Both analyses hence agree on inferring a very small amount of heterogeneity (mostly below $\tau=0.1$), and consistency with homogeneity ($\tau=0$).

    \subsubsection{Random heterogeneity parameters}  
      Instead of a single common heterogeneity parameter, we may also use a random-effects model for the heterogeneity and infer a \emph{distribution} of heterogeneity parameters. Such an approach was proposed by \citet{RhodesEtAl2015} and \citet{TurnerEtAl2015}; here we are utilizing the implementation described by \citet{RoeverEtAl2023}.
      Specifying a common half-normal distribution for the 13~heterogeneity parameters~$\tau_j$, and a uniform hyperprior (with a ``large'' upper bound of~10) for the associated scale parameter~$\varphi$,
      \begin{equation}
        \tau_j \sim \halfnormaldistn(\varphi)
        \quad \mbox{and} \quad
        \varphi \sim \unifdistn(0,\,10) \mbox{,}
      \end{equation}
      we yield a (predictive) distribution with a median of~$0.035$ and an upper 95\% quantile of~$0.22$ for the heterogeneity~$\tau^\star$.
      Again, the analysis points to a very small amount of heterogeneity, when considering the  (logarithmic) effect scale investigated here \citep{RoeverEtAl2021}.


\section{Reconsidering the two examples}
  \subsection{The GLOW trials}\label{sec:glow2}
    Heterogeneity estimates for the \textsc{Glow~1/2} data (see Section~\ref{sec:glow1}) are shown in Table~\ref{tab:GlowHeterogeneity} based on several analysis methods.
    Cochran's $Q$-test for homogeneity yields a $p$-value of~0.62 here (i.e., no indication of heterogeneity), and the frequentist point estimate of heterogeneity is also at $\hat{\tau}=0$.
    Note that for $k=2$ studies, the most common heterogeneity estimators, like DerSimonian-Laird (DL), restricted maximum likelihood (REML) or Paule-Mandel (PM), all coincide \citep{Rukhin2012,FriedeRoeverWandelNeuenschwander2017b}.
      \begin{table} 
        \caption{\label{tab:GlowHeterogeneity}Estimated heterogeneity based on the \textsc{Glow~1/2} data from Figure~\ref{fig:glow} and several random-effects methods. Frequentist estimates are identical for the most common methods, and frequentist CIs are based on the $Q$-profile method \citep{Viechtbauer2007}.  The a-priori expected heterogeneity 
        for the half-normal HN(10) prior it is 6.7 [0.0, 19.6],
        for the HN(20) prior is 13.5 [0.0, 39.2],
        and for the HN(50) prior is 33.7 [0.0, 98.0].}
        \centering
        \begin{tabular}{lcc}
          \toprule
            method           & $\hat{\tau}$ & 95\% CI \\
          \midrule
            frequentist (DL/REML/PM)       & \phantom{0}0.0 & [0.0, 247.7] \\ 
            Bayes (HN(10))   & \phantom{0}6.1 & [0.0, 18.1] \\ 
            Bayes (HN(20))   & 10.5 & [0.0, 33.4] \\ 
            Bayes (HN(50))   & 19.4 & [0.0, 73.9] \\ 
          \bottomrule
        \end{tabular}
      \end{table}
    A Bayesian analysis of the data requires specification of a (proper, informative) heterogeneity prior that is suitable in the context of the present endpoint (the FEV\textsubscript{1} mean difference).
    The FEV\textsubscript{1} itself is of the order of 1.3--1.4~L\@ at inclusion, with standard deviations of 0.5~L among patients \citep{DUrzoEtAl2011,KerwinEtAl2012}.
    Considering the \textsc{Glow} trials' standard errors and sample sizes, the \emph{unit information standard deviation (UISD)} (for the FEV\textsubscript{1} pre/post \emph{difference}) amounts to~$\sigma_\mathrm{u}=446$~mL for the example data \citep{RoeverEtAl2021}.
    In a related review investigating a different bronchidilator (Tioptropium) in a similar context, \citet{KarnerChongPoole2014} observed a heterogeneity of $\tau\approx20$~mL among a set of~22 (most likely more heterogeneous) included studies.
    Table~\ref{tab:GlowHeterogeneity} hence lists Bayesian estimates based on half-normal priors with scale parameters of $10$, $20$ and $50$~mL.

      \begin{table} 
        \caption{\label{tab:GlowEstimates}Effect estimates (mean difference) based on the \textsc{Glow~1/2} data from Figure~\ref{fig:glow} and on the common-effect (fixed effect, FE) and several random-effects (RE) methods. The Hartung-Knapp-Sidik-Jonkman (HKSJ) and modified Knapp-Hartung (mKH) methods are based on adjusted standard errors and Student\mbox{-}$t$ quantiles, the Bayesian estimates are based on half-normal (HN) heterogeneity priors of differing scales.} \centering
        \begin{tabular}{lccc}
          \toprule
            method & MD & 95\% CI & CI width\\
          \midrule
            FE    & 103.2 & [81.4, 124.9] & \phantom{0}43.5 \\ 
            RE    & 103.2 & [81.4, 124.9] & \phantom{0}43.5 \\ 
            HKSJ  & 103.2 & [33.8, 172.5] & 138.7 \\ 
            mKH   & 103.2 & [--37.7, 244.0] & 281.7 \\ 
            Bayes (HN(10)) & 103.1 & [77.7, 128.3] & \phantom{0}50.6 \\ 
            Bayes (HN(20)) & 103.0 & [70.6, 135.1] & \phantom{0}64.5 \\ 
            Bayes (HN(50)) & 102.9 & [45.7, 159.5] & 113.8 \\ 
          \bottomrule
        \end{tabular}
      \end{table}

    Table~\ref{tab:GlowEstimates} shows several estimates of the mean effect along with 95\% CIs. Due to the heterogeneity being estimated as zero, the fixed-effect (FE) and random-effects (RE) estimates are identical. The Hartung-Knapp-Sidik-Jonkman (HKSJ) interval is wider, while the modified Knapp-Hartung (mKH) interval renders the overall effect not significant.
    This behaviour is to some extent expected --- according to the taxonomy of \citet{JacksonEtAl2017}, here we are in a case where the data appear homogeneous (with a zero heterogeneity estimate), so that the difference between the RE and mKH intervals is only the use of a Student-$t$ quantile (with 1~degree of freedom) rather than the corresponding standard normal quantile, so that the mKH~interval will end up about 6.5~times as wide.
    The different Bayesian analyses yield wider intervals than FE, but all of them remain shorter than the HKSJ interval.
    \citet{DUrzoEtAl2013} later quoted a pooled estimate of
    $103.0$ [$81.0$, $125.0$]
    based an analysis of the pooled data (stratified by study), which roughly matches the FE estimate.

\subsection{The RESPIRE trials}\label{sec:respire2}
  Point estimates and significance tests for heterogeneity among the logarithmic incidence rate ratios (IRRs) in the \textsc{Respire} example data (see Section~\ref{sec:respire1}) yield $\hat{\tau}=0.13$ ($p=0.22$) for the 14-day regimen, and $\hat{\tau}=0.36$ ($p=0.028$) at 28 days.
  The $Q$-test outcome is significant at the 5\%~level only for the 28-day regimen.
  A Bayesian meta-analysis again requires a proper, informative prior. For the present endpoint on a logarithmic scale, a half-normal prior with scale~$0.50$ is a reasonable choice; given the relatively small expected heterogeneity among a pair of study twins, a smaller scale (e.g., $0.25$) may also be considered appropriate \citep{RoeverEtAl2021}. The corresponding frequentist and Bayesian heterogeneity estimates are shown in Table~\ref{tab:RespireHeterogeneity}.

  The heterogeneity estimates are not extremely large; values up to $0.5$ are commonly regarded as ``reasonable'' \citep{SpiegelhalterEtAl,RoeverEtAl2021}, however, in this particular case of a twin pair of studies, the outcomes might be expected to be rather homogeneous. A heterogeneity value of~$\tau=0.36$ would imply that study-specific IRRs vary around a common mean value by factors roughly within~$0.49$ and~$2.0$ (with 95\% probability) \citep{RoeverEtAl2021}.
  When considering the Bayesian estimates and 95\% intervals, it becomes clear that the posteriors differ little from the assumed priors, i.e., the data provide rather little evidence on the amount of heterogeneity actually present.

  We may also compare the observed (estimated) heterogeneity magnitude to what was observed in the NEJM sample when we consider the subgroup of study pairs that also investigated endpoints on a logarithmic scale.  Among the 13 ``logarithmic'' endpoints in NEJM sample (log-OR, log-RR, log-IRR or log-HR endpoints), 8~heterogeneity estimates turned out as zero; we have mean $\hat{\tau}$ of 0.052 and a maximum of 0.27; so the \textsc{Respire} studies' estimate of $\hat{\tau}=0.36$ is indeed at the upper end of the observed range in this (albeit small) sample.

      \begin{table} 
        \caption{\label{tab:RespireHeterogeneity}Estimated
          heterogeneity based on the \textsc{Respire~1/2} data from
          Figure~\ref{fig:respire} and several random-effects
          methods (corresponding to the \emph{logarithmic} IRR). Frequentist estimates are identical for several
          methods, and frequentist CIs are based on the $Q$-profile
          method \citep{Viechtbauer2007}.  The a-priori expected
          heterogeneity for the half-normal HN(0.25) prior is 0.17 [0.00, 0.49],
          and for the HN(0.50) prior it is 0.34 [0.00, 0.98].}
        \centering
        \begin{tabular}{llcc}
          \toprule
            regimen & method & $\hat{\tau}$ & 95\% CI \\
          \midrule
            14-day   & frequentist (DL/REML/PM) & 0.13 & [0.00, 6.98] \\ 
                     & Bayes (HN(0.25))   & 0.15 & [0.00, 0.49] \\ 
                     & Bayes (HN(0.50))   & 0.25 & [0.00, 0.80] \\ 
          \cmidrule(lr){1-4}
            28-day   & frequentist (DL/REML/PM) & 0.36 & [0.00, 13.1] \\ 
                     & Bayes (HN(0.25))   & 0.23 & [0.00, 0.51] \\ 
                     & Bayes (HN(0.50))   & 0.37 & [0.00, 0.92] \\ 
          \bottomrule
        \end{tabular}
      \end{table}

    Table~\ref{tab:RespireEstimates} shows effect estimates from common-effect as well as several random-effects analyses of the \textsc{Respire~1/2} data.
      \begin{table} 
        \caption{\label{tab:RespireEstimates}\textsc{Respire~1/2} effect estimates (incidence rate ratios) based on the data from Figure~\ref{fig:respire} and on the common-effect (fixed effect, FE) and several random-effects (RE) methods.}  \centering
        \begin{tabular}{llccc}
          \toprule
            regimen & method & IRR & 95\% CI & CI width\\
          \midrule
            14-day & FE    & 0.718 & [0.561, 0.919] & 0.359 \\ 
                   & RE    & 0.716 & [0.529, 0.970] & 0.441 \\ 
                   & HKSJ  & 0.716 & [0.100, 5.110] & 5.010 \\ 
                   & mKH   & 0.716 & [0.100, 5.110] & 5.010 \\ 
                   & Bayes (HN(0.50)) & 0.715 & [0.373, 1.362] & 0.988 \\ 
                   & Bayes (HN(0.25)) & 0.716 & [0.474, 1.079] & 0.605 \\ 
          \cmidrule(lr){1-5}
            28-day & FE    & 0.732 & [0.566, 0.949] & 0.383 \\ 
                   & RE    & 0.733 & [0.416, 1.294] & 0.878 \\ 
                   & HKSJ  & 0.733 & [0.019, 29.02] & 29.00 \\ 
                   & mKH   & 0.733 & [0.019, 29.02] & 29.00 \\ 
                   & Bayes (HN(0.50)) & 0.733 & [0.337, 1.599] & 1.262 \\ 
                   & Bayes (HN(0.25)) & 0.733 & [0.448, 1.120] & 0.751 \\ 
          \bottomrule
        \end{tabular}
      \end{table}
    The analysis methods used were a common-effect (or fixed-effect (FE)) analysis, a random-effects (RE) analysis utilizing a plug-in estimate for the heterogeneity variance and confidence intervals based on a normal approximation, Hartung-Knapp-Sidik-Jonkman (HKSJ) and modified Knapp-Hartung (mKH) intervals based on a Student\mbox{-}$t$ approximation \citep{RoeverKnappFriede2015}, and a Bayesian analysis utilizing a half-Normal prior with scale~$0.50$ or~$0.25$ for the heterogeneity standard deviation \citep{Roever2020,RoeverEtAl2021}.
    Whether the evidence on the treatment effect is deemed conclusive ($\mbox{IRR}<0$), in this case depends on the model assumptions implemented.
    With positive $\tau$~estimates suggesting the presence of heterogeneity (Table~\ref{tab:RespireHeterogeneity}), the HKSJ and mKH intervals are both substantially wider than the simple RE interval. In this case, the width inflation (on the \emph{logaritmic} IRR scale) is by a factor of~6.5 for both HKSJ and mKH \citep{JacksonEtAl2017}. While the Bayesian intervals also account for the uncertainty in~$\tau$, the corresponding intervals are much closer to the RE~interval.

\section{Discussion}
  The present investigations showed that in general a single pair of effect estimates provides only very little evidence on the heterogeneity between studies. Even if studies are in fact homogeneous, they are not unlikely to suggest heterogeneity, or, conversely, even under substantial heterogeneity, the data may still appear homogeneous. Ad-hoc criteria employed to judge heterogeneity will often be misleading.
  When considering a larger number of published pairs of ``study twins'', overall these appear rather homogeneous, but it remains unclear whether this might be due to the fact that clinical and methodological homogeneity (similarity in study designs and protocols) in fact does lead to statistical homogeneity (consistent effect estimates), or whether this observation may be due to selection effects. Apparent heterogeneity might have triggered decisions precluding a joint analysis or its publication in a high-ranking journal such as the NEJM\@.
  In order to overcome this possible limitation, we are currently considering ways of investigating pairs of study twins without the restriction to a joint publication.

  In the presence of (known or potential) effect modifiers, a certain amount of heterogeneity should be anticipated \citep{Higgins2008}; use of meta-regression to adjust for factors that may not be controlled by the study design then might help reconciling cases like the \textsc{Respire} example of (apparent) clinical and statistical heterogeneity despite methodological homogeneity. However, the danger of overfitting based on a very small number of studies needs to be considered \citep{HartungKnappSinha}.
  Careful reporting of studies' relevant characteristics may then help interpreting heterogeneous study results, and may facilitate valuable insights.
  On the other hand, efforts to ensure methodological homogeneity (standardization of protocols etc.) and measures of clinical homogeneity (patient characteristics) would need to be transparently and convincingly documented in order to possibly justify assuming little or no heterogeneity in a meta-analysis.
  The amount of residual heterogeneity to be expected may also depend on the indication in question, or the degree to which the underlying biology is understood \citep{RhodesEtAl2015,TurnerEtAl2015}. While we did not see clear evidence of heterogeneity in our example data, possible future investigations based on larger data sets may need to consider distinct categories of applications.

  Heterogeneous results clearly do occur in practice \citep{KontopantelisSpringateReeves2013,RhodesEtAl2015,TurnerEtAl2015}.
  While heterogeneity may be perceived as a threat to external validity \citep{DiasEtAl2013c}, 
  \citet{IoannidisTrikalinosZintzaras2006} in fact warn that ``extreme homogeneity'' may also raise suspicion, and 
  \citet{Higgins2008} questions the external validity of a meta-analysis when studies have been ``sieved'' for homogeneity. Efforts to ensure methodological and clinical homogeneity may on the one hand promote homogeneous and concordant outcomes, but might to some extent also counteract external validity or generalizability.
  This also points to the related question of whether the conduct of several (potentially heterogeneous) trials might sometimes be preferrable to a single, larger trial \citep{ShrierPlattSteele2007}, but, besides obvious budget constraints, this will also depend on whether the actual research question is of a confirmatory or exploratory nature \citep{Tukey1980}.
  With meta-analyses of study twins, the common-effect model is often a sensible option. Sensitivity analyses might include Bayesian random-effect meta-analyses with prior distributions for the heterogeneity placing substantial probability at small levels of heterogeneity.


\section*{Acknowledgment}
  Support from the \emph{Deutsche Forschungsgemeinschaft (DFG)} is
  gratefully acknowledged (grant number \mbox{FR~3070/3-1}).

\section*{Conflicts of interest}
  The authors have declared no conflict of interest.


\begin{appendix}
  \section{Appendix}
  \subsection{The difference in estimates}
    In the special case of only $k=2$ studies, the homogeneity or heterogeneity of studies only manifests itself in the (absolute) \emph{difference} between the two estimates involved, $|y_2-y_1|$.  According to the model assumptions, the estimates' difference follows a normal distribution:
    \begin{equation}
      (y_2-y_1) \;\sim\; \normaldistn(0,\, \sigma_1^2+\sigma_2^2+2\tau^2)
      \mbox{.}
    \end{equation}
    The implications for the \emph{absolute} difference may then be investigated by noting that
    \begin{equation}\label{eqn:chi2distn}
      \frac{(y_2-y_1)^2}{\sigma_1^2+\sigma_2^2+2\tau^2} \;\sim\; \chi^2_1 \mbox{,}
    \end{equation}
    i.e., the squared (and scaled) difference follows a $\chi^2$~distribution with one degree of freedom. Probabilities of exceeding certain thresholds may hence be computed via
    \begin{equation}
      \prob\bigl((y_2-y_1)^2\leq c^2 \,\big|\, \sigma_1, \sigma_2, \tau\bigr)
      \;=\;
      F_{\chi^2_1}\Bigl(\textstyle\frac{c^2}{\sigma_1^2+\sigma_2^2+2\tau^2}\Bigr)
      \mbox{,}
    \end{equation}
    where $F_{\chi^2_1}(\cdot)$ denotes the $\chi^2_1$~distribution's cumulative distribution function (CDF). In the special case of equal standard errors ($\sigma_2=\sigma_1=\sigma$), the expression simplifies further.
    In the following subsections, we will consider several thresholds for the difference in estimates that may often be taken as an indication of statistical homogeneity or heterogeneity.

  \subsection{Relevant thresholds for the difference}
  \subsubsection{Overlapping CIs}
    Two studies exhibit overlapping 95\%~CIs whenever
    \begin{equation}\label{eqn:overlapCondition}
      |y_2-y_1| \;\leq\; z_{0.975}(\sigma_1 + \sigma_2)\mbox{,}
    \end{equation}
    where $z_{0.975}\approx1.96$ is the 97.5\% quantile of the standard normal distribution.

  \subsubsection{Mutual estimate coverage}
    A somewhat more strict condition is that each of the two CIs contains the other effect estimate. This happens whenever
    \begin{equation}\label{eqn:mutualCondition}
      |y_2-y_1| \;\leq\; z_{0.975}\, \min\{\sigma_1,\, \sigma_2\} \mbox{.}
    \end{equation}

  \subsubsection{A zero heterogeneity estimate}
    Frequentist heterogeneity estimates frequently turn out as zero
    \citep{FriedeRoeverWandelNeuenschwander2017a}.
    In the case of only $k=2$~studies, this happens for the more common estimators (at least for the DerSimonian-Laird, maxi\-mum-like\-li\-hood, restricted maxi\-mum-like\-li\-hood and Paule-Mandel estimators) whenever
    \begin{equation}\label{eqn:zeroTauCondition}
      (y_2-y_1)^2 \;\leq\; \sigma_1^2+\sigma_2^2 
    \end{equation}
    \citep{Rukhin2012,FriedeRoeverWandelNeuenschwander2017b}.

  \subsubsection{A non-significant $Q$~test}\label{sec:CochranQ}
    Cochran's $Q$~test is a statistical test for the null hypothesis of \emph{homogeneity} ($\tau=0$). The test statistic is
    \begin{equation}
      Q \;=\; \sum_{i=1}^k \frac{(y_i-\hat{\mu})^2}{\sigma_i^2}
      \mbox{,}\qquad\mbox{where}\qquad
      \hat{\mu} \;=\; \frac{\sum_{i=1}^k \frac{y_i}{\sigma_i^2}}{\sum_{i=1}^k \frac{1}{\sigma_i^2}}
      \mbox{.}
    \end{equation}
    The null hypothesis is rejected if 
    $Q > \chi^2_{(k-1); (1-\alpha)}$,
    where $\alpha$ is the type-I error allowed for
    \citep{Cochran1954,Fleiss1993}.
    For the case of only two studies ($k=2$), we have
    \begin{equation}\label{eqn:Qk2}
      Q \;=\; \frac{(y_2 - y_1)^2}{\sigma_1^2 + \sigma_2^2}
      \mbox{,}
    \end{equation}
    so that the $Q$~test rejects if
    \begin{equation}
      (y_2 - y_1)^2 \;>\; \chi^2_{1;(1-\alpha)}  (\sigma_1^2 + \sigma_2^2)
      \mbox{,}
    \end{equation}
    where in the common case of $\alpha=5\%$, we have
    $\chi^2_{1;0.95}=(z_{0.975})^2=3.84$.

  \subsection{Probabilities of threshold exceedance}
    Based on the considerations from the previous subsections, we can derive probabilities for certain events that may commonly be taken as indicative of (``statistical'') homogeneity.
    Table~\ref{tab:empiricalHeteroProb} lists thresholds for the difference in estimates ($|y_2-y_1|$) as well as the expressions to compute the probability of remaining below that threshold. For the case of equal standard errors ($\sigma_2=\sigma_1=\sigma$), these expressions simplify. In that case, there is also a one-to-one correspondence of~$\tau$ to the popular $I^2$~heterogeneity measure proposed by \citet{HigginsThompson2002}, which results as $I^2=\frac{\tau^2}{\tau^2 + \sigma^2}$ (so that
    $\frac{\sigma^2}{\sigma^2+\tau^2}=(1-I^2)$).

    \begin{table}[ht] 
      \caption{\label{tab:empiricalHeteroProb}Probabilities for events that may be seen as an indication of homogeneity. Each of the events is present whenever the (absolute) difference in estimates $|y_2-y_1|$ is below a certain threshold. Event probabilities may then be computed based on the CDF of a $\chi^2$-distribution with 1~degree of freedom ($F_{\chi^2_1}$). $z_{0.975}\approx 1.96$~here denotes the 97.5\%-quantile of a standard normal distribution, and $\alpha$~is the significance level of the $Q$-test.}  \centering
      \begin{tabular}{lcccc}
        \toprule
              & threshold & \multicolumn{2}{c}{probability} \\
        \cmidrule(lr){3-4}
        event & for $|y_2-y_1|$ & general case & special case \\
              & & $\sigma_2\neq\sigma_1$ & $\sigma_2=\sigma_1=\sigma$\\
        \midrule
        CI overlap         & $z_{0.975} \, (\sigma_1 + \sigma_2)$ & $F_{\chi^2_1}\Bigl(\textstyle\frac{z_{0.975}^2\,(\sigma_1 + \sigma_2)^2}{\sigma_1^2+\sigma_2^2+2\tau^2}\Bigr)$ & $F_{\chi^2_1}\Bigl(\textstyle 2 z_{0.975}^2 \,\frac{\sigma^2}{\sigma^2+\tau^2}\Bigr)$\\
        non-sig.\ $Q$-test & $z_{(1-\alpha/2)} \, \sqrt{\sigma_1^2 + \sigma_2^2}$ & $F_{\chi^2_1}\Bigl(\textstyle\frac{z_{(1-\alpha/2)}^2\,(\sigma_1^2 + \sigma_2^2)}{\sigma_1^2+\sigma_2^2+2\tau^2}\Bigr)$ & $F_{\chi^2_1}\Bigl(\textstyle z_{(1-\alpha/2)}^2 \, \frac{\sigma^2}{\sigma^2+\tau^2}\Bigr)$\\
        mutual coverage    & $z_{0.975}\, \min\{\sigma_1,\, \sigma_2\}$ & $F_{\chi^2_1}\Bigl(\textstyle\frac{z_{0.975}^2(\min\{\sigma_1,\, \sigma_2\})^2}{\sigma_1^2+\sigma_2^2+2\tau^2}\Bigr)$ & $F_{\chi^2_1}\Bigl(\textstyle\frac{z_{0.975}^2}{2}\frac{\sigma^2}{\sigma^2+\tau^2}\Bigr)$\\
        $\hat{\tau}=0$     & $\sqrt{\sigma_1^2+\sigma_2^2}$ & $F_{\chi^2_1}\Bigl(\textstyle\frac{\sigma_1^2+\sigma_2^2}{\sigma_1^2+\sigma_2^2+2\tau^2}\Bigr)$ & $F_{\chi^2_1}\Bigl(\textstyle\frac{\sigma^2}{\sigma^2+\tau^2}\Bigr)$\\
        \bottomrule
      \end{tabular}
    \end{table}

  \subsection{Distribution of $Q$ under alternatives ($H_1$, $\tau>0$)}
    Figure~\ref{fig:pQvalues} in Section~\ref{sec:descriptive} showed the CDF of the $Q$-test statistic; the computation of this distribution is given in the following.
    Since for $k=2$ the test statistic simplifies to the expression in~(\ref{eqn:Qk2}), and the distribution of the difference in estimates is known to follow a (scaled) $\chi^2$-distribution (\ref{eqn:chi2distn}), the $Q$-statistic's distribution under an alternative simply follows a scaled $\chi^2$-distribution. For example, the cumulative distribution function~$F_\tau$ under an alternative ($\tau>0$) is given by
    \begin{equation}
      F_\tau(x) \;=\; F_{\chi^2_1}\Bigl(x \textstyle \frac{\sigma_1^2+\sigma_2^2}{\sigma_1^1+\sigma_2^2+2\tau^2}\Bigr)\mbox{.}
    \end{equation}
    Note that for the case of identical standard errors ($\sigma_2=\sigma_1=\sigma$), the factor again simplifies to
    \begin{equation}
      \frac{\sigma_1^2+\sigma_2^2}{\sigma_1^1+\sigma_2^2+2\tau^2}
      \;=\;
      \frac{2\sigma^2}{2\sigma^2+2\tau^2}
      \;=\;
      \frac{\sigma^2}{\sigma^2+\tau^2}
      \;=\;
      1-I^2 \mbox{.}
    \end{equation}

\end{appendix}

\bibliographystyle{bimj}
\bibliography{literature}

\end{document}